\documentclass{article}
\begin{document}
\font\addressfont = cmti10
\title{Perihelion Precession in the Special Relativistic\\ Two-Body Problem}
\author{M.~A. Trump and W.~C. Schieve\\
\mbox{}\\
\addressfont
Department of Physics and Ilya Prigogine Center for Studies in\\ 
\addressfont 
Statistical Mechanics and Complex Systems,\\
\addressfont
The University of Texas at Austin, Austin, Texas, 78712}
\date{\today}
\maketitle
{\noindent {\em Abstract\/}: The classical two-body system  
with Lorentz-invariant Coulomb action-at-a-distance $V=-k/\rho$
is solved in 3+1 dimensions using the manifestly covariant 
Hamiltonian
mechanics of St\"{u}ckelberg.
Particular solutions for the reduced motion are obtained which
correspond to
bound attractive, unbound attractive,
and repulsive scattering motion. 
A lack of perihelion precession is found
in the bound attractive orbit, and the
semi-classical hydrogen spectrum subsequently contains
no fine structure corrections. It is argued that this
prediction is indicative of the correct classical 
special relativistic two-body theory.}

\section{Introduction}

One of the most significant applications 
of the St\"{u}ckelberg relativistic
dynamics \cite{Stueckelberg1941,HorwitzPiron1973}
is certainly the classical two-body problem.
The significance arises from
the fact that the St\"{u}ckelberg mechanics 
allows the use of {\em Lorentz-invariant action-at-a-distance} 
in the form $V=V(\rho)$, $\rho=\rho(\tau)$\footnote{The parameter $\tau$ is the
invariant dynamical time; see ref. \cite{HorwitzPiron1973}.}, to 
model long-range mutual interaction between particles.

In this paper 
we study 
the classical covariant Kepler problem, i.e.,
the classical special relativistic two-body problem
with scalar Coulomb action-at-a-distance $V=-k/\rho$, 
where $k>0$ is an invariant constant. 
Although certain solutions for this potential were obtained
by Cook \cite{Cook1972}
and by Piron and Reuse \cite{PironReuse1975}, 
the full range of physical motion 
produced by this potential 
has not yet been investigated.

In this paper 
we outline\footnote{A more complete discussion
        of these solutions is given in ref.~\cite{TrumpSchieveBook}. The
        1+1-dimensional version of this system is also discussed in
         ref.~\cite{TrumpSchieve1998}.} 
two separate methods 
to study 
the reduced two-body motion 
for this potential 
using the pseudospherical coordinates $(\rho,\beta,\theta,\phi)$
suggested  by Cook\footnote{In ref.~\cite{Cook1972}; 
                   see Section~3 below for a discussion of these coordinates.}.
Both methods are 
generalizations of solutions 
of the nonrelativistic inverse-square potential $V=-k/r$.
In the first method, 
we integrate 
the Euler-Lagrange equations
for the central force $V=V(\rho)$ 
to obtain
a set of Lorentz-invariant isolating integrals, 
which are solved
for a set of classical {\em orbit equations} 
in the pseudospherical coordinates.
We plot 
particular solutions 
for the invariant Coulomb potential
in the 2+1 dimensional 
particle separation coordinates
$(\Delta x (\tau), \Delta y (\tau), \Delta t(\tau))$.
In the second method, 
we solve 
the classical Hamilton-Jacobi equation  
to obtain 
the Lorentz-invariant {\em action variables}
for the central force potential.
In the case of the Coulomb potential,
we impose 
the rule of semi-classical quantization
to obtain a prediction 
for the electron energy levels 
of hydrogen. 

The classical orbit solutions 
of the potential $V=-k/\rho$
will be shown to produce 
all three types of physical motion
expected for 
the two-body system 
with inverse-square interaction---i.e.,
bound attractive, 
unbound attractive, 
and repulsive scattering motion---with 
the distinction 
being given 
in a straightforward manner 
by a pair of 
Lorentz-invariant eccentricity constants\footnote{Here we set $k>0$ 
           and obtain both attractive and repulsive solutions. 
           The distinction 
           between the two cases is absorbed in the definition of the 
           eccentricity constants; see ref.~\cite{TrumpSchieveBook}.}.
The most striking prediction 
in these solutions 
is the lack of 
perihelion precession 
in the bound orbits 
of the special relativistic system\footnote{cf. A.~Sommerfeld, 
          1915 \cite{Sommerfeld1915}}.
As a result of 
this lack of precession, 
the semi-classical hydrogen spectrum 
contains no fine structure corrections.
It is argued that 
this result 
is favorable 
to the St\"{u}ckelberg theory,
based on consideration 
of the classical limit 
without spin.

\section{The Classical and Semi-Classical Solutions}

Consider 
an isolated system 
of two particles 
with constant finite rest masses $m_i$, $i=1,2$, 
and mutual long-range interaction 
over flat spacetime.
It is well known \cite{HorwitzPiron1973,Fanchi,HorwitzSchieve1992,TrumpSchieve1997} 
that in the covariant mechanics,
the most straightforward solution 
of the classical problem 
with scalar interaction
is obtained 
by a transformation 
from the individual particle 
event coordinates
$(x^\mu_1(\tau),x^\mu_2(\tau))$, 
$\mu=0,1,2,3$,
to the
covariant {\em center-of-mass and reduced coordinates},
$(X^\mu(\tau), x^\mu(\tau))$.
The system can be regarded 
as comprising two fictitious particles 
with masses $M$ and $m$. The transformations are 
\begin{eqnarray}
M &=& m_1 + m_2, \qquad m=\left({m_1 \,m_2}\right)/M,\label{Mm}\\
X^\mu &=& \left({m_1 x^\mu_1 + m_2 x^\mu_2}\right)/M,\qquad 
x^\mu = x^\mu_2 - x^\mu_1
\label{Xx}
\end{eqnarray}
In these coordinates,
the nontrivial motion of the system 
occurs entirely in the reduced problem, 
for which the frame components are
$x^\mu=(\Delta t,\Delta x, \Delta y,\Delta z)$. 

The most convenient procedure 
is to make a second coordinate transformation 
of the reduced problem 
to the pseudospherical coordinates of Cook \cite{Cook1972},
which consist of the variable spacetime separation $\rho=\rho(\tau)$, 
as well as a spacetime angle $\beta=\beta(\tau)$, and two ordinary
angles $\theta=\theta(\tau)$ and $\phi=\phi(\tau)$. The transformations are
\begin{eqnarray}
\rho 
&=&\left[{ \left({\Delta x}\right)^2 + \left({\Delta y}\right)^2
+\left({\Delta z}\right)^2 -\left({\Delta t}\right)^2
}\right]^{1/2}=\left|{x^\mu}\right|\nonumber\\
\tanh \beta&=&
{{\Delta t}
/{\sqrt{\left({\Delta x}\right)^2 + 
\left({\Delta y}\right)^2 
+\left({\Delta z}\right)^2}}}\label{rho}\\
\tan \theta&=&
 {{\Delta z}/{\sqrt{\Delta x^2+\Delta y^2}}}
\nonumber\\
\tan \phi&=& {{\Delta y}/{\Delta x}}.
\nonumber
\end{eqnarray}

In these coordinates,
the Lagrangian 
for a central force potential is
\begin{equation}
L={1 \over 2}\,m\left({
\dot{\rho}^2-\rho^2\,\dot{\beta}^2+\rho^2\cosh^2\beta\, \dot{\theta}^2
+\rho^2\cosh^2\beta \sin^2\theta\,\dot{\phi}^2
}\right) -V(\rho).
\label{lagrangian}
\end{equation}
where the dot indicates 
differentiation 
with respect to the world time.
By integrating 
the four resultant 
Euler-Lagrange equations 
in the pseudospherical coordinates $(\rho,\beta,\theta,\phi)$, 
one obtains
the Lorentz-scalar isolating integrals, 
\begin{eqnarray}
l_{\phi}&=&m\,\rho^2 \cosh \beta \dot{\phi},\nonumber\\
l_{\theta}&=&\left[{m^2\,\rho^4 
     \cosh^2 \beta \dot \theta^2 +  
    l_{\phi}^2\,\csc^2 \theta}\right]^{1/2},\label{integrals}\\
\Lambda^2&=& m^2\rho ^4\dot \beta ^2-l_\theta ^2\,\mbox{sech}^2\beta,\nonumber\\
E&=&{1 \over 2}m\dot \rho ^2+{{\Lambda ^2} \over {2m\rho ^2}}
				+V\left( \rho  \right).\nonumber
\end{eqnarray}
It can be shown\cite{HorwitzSchieve1992} 
that the conservation 
of a covariant angular momentum tensor
implies that the polar angle
$\theta$ can be eliminated from the problem by an ordinary
three-rotation of the coordinate axes, i.e., by choice of a frame
in which $\dot{\theta}\equiv 0$ and $\theta\equiv \pi/2$.
Since this rotation does not involve relativity, it then possible to 
study the classical problem in 2+1 dimensions without loss of generality.

By comparing differential expressions, the three remaining 
integrals of the motion $(l_{\phi}, \Lambda^2, E)$ may be solved to
derive a pair of central force {\em orbit 
equations} \cite{TrumpSchieveBook} for the dependence of
$\rho$ and $\phi$ upon the spacetime angle $\beta$.
The azimuthal orbital component  $\phi=\phi(\beta)$ is given by
\begin{equation}
\phi(\beta)= \tan^{-1} \left({
{{q\sinh(\beta-\beta_0)}\over{\sqrt{q^2 - \cosh^2 (\beta-\beta_0)}}}
}\right),
\label{phiorbit}
\end{equation}
where $q^2\equiv -l_{\phi}^2/\Lambda^2$.
For a conservative potential $V=V(\rho)$,
the radial orbital component $\rho=\rho(\beta)$ is given
by the solution
of the differential equation,
\begin{equation}
\left( {1-{{q^2} \over {\cosh ^2\beta }}} \right)
{{d^2 u }\over{d \beta^2}}+q^2{{\tanh \beta } \over {\cosh ^2\beta }}
{{du}\over{d\beta}}
-u=+{m \over 
{\Lambda ^2}}{{dV} \over {du}},
\label{rhoorbit}
\end{equation}
where $u\equiv 1/\rho$.

In the case of the {\em two-body Coulomb potential},
\begin{equation}
V= {{-k}\over{\rho}},\qquad k=\mbox{Lorentz-invariant constant},
\end{equation}
the inhomogeneous term
in the radial orbit equation (\ref{rhoorbit}) is a constant,
and thus
particular solutions are known trivially once the homogeneous
equation is solved.
It is useful to examine the two particular solutions formed
by adding the inhomogeneous term to the two linearly independent solutions
of the homogeneous equation, which may be labeled {\em Type~I} and
{\em Type~II solutions} for $\rho=\rho(\beta)$,
\begin{eqnarray}
\mbox{Type I:}\qquad{1 \over {\rho}}&=&{{mk} \over {\Lambda ^2}}
\left({
1-{{e'}\over{f}}\,\sinh \left({\beta-\beta_0}\right) }\right),
\label{TypeI}
\\
\mbox{Type II:}\qquad
{1 \over {\rho}}&=&{{mk} \over {\Lambda ^2}}
\left({
1-{{e''}\over{f}}\,\left[{
{\left( {\sinh ^2\left( {\beta -\beta _0} 
\right)-f^2} \right)^{{1 \mathord{\left/ {\vphantom {1 2}} \right. 
\kern-\nulldelimiterspace} 2}}}
}\right]}\right),
\label{TypeII}
\end{eqnarray}
where $f\equiv \sqrt{q^2-1}$, and where
$e'$ are $e''$ are two new invariant constants which are 
explicit functions of
$E$ and $\Lambda$ \cite{TrumpSchieveBook}. 
They may be regarded as Lorentz-invariant generalizations
of the Kepler eccentricity \cite{Goldstein}.
Note that
$f$ is the same constant in both (\ref{TypeI}) and (\ref{TypeII}).

From the orbital solutions $\theta=\theta(\beta)$ and $\rho=\rho(\beta)$,
the physical motion produced by the Coulomb interaction can
be understood by using the inverse transformations of (\ref{rho})
to plot in the frame coordinates 
$(\Delta x(\beta), \Delta y(\beta), \Delta t(\beta))$.
If it is desired,
the $\tau$ dependence of the solution may be included 
by numerical inversion of
the expression $\tau=\tau(\beta)$ obtained from
integration of eqs.~(\ref{integrals}) \cite{TrumpSchieveBook}.

It is useful to examine the dependence of the Type I and Type II solutions
upon the eccentricities $e'$ and $e''$, as well
as upon $f$, 
the rotational constant\footnote{So-called because it 
                    depends explicitly on $l_{\phi}$ in 2+1 dimensions.}.
For a given $e'$ or $e''$, the effect of
the variation of $f$ is simply to modify the scale and proportion of
the curve (see \cite{TrumpSchieveBook}).
The topology of the reduced orbit---and thus the classification
of the physical motion---is found to depend on the magnitude of
$e'$ or $e''$.

Fig.~(\ref{FigOne}) is a summary of the classification of orbits
arising from variation of $e'$ and $e''$. For a proof of the
properties discussed here, see ref.~\cite{TrumpSchieveBook}.
From the three
panels of the diagram, it may be seen that the Type~I
solutions correspond to an attractive Coulomb interaction, whereas
Type~II corresponds to a repulsive potential. For the Type~I solutions,
the effect of the transformation $e' \to -e'$ is a reflection of the
curve across the $(\Delta x,\Delta y)$ plane, and thus one need consider only
$\left|{e'}\right|$. For the range 
$0<e'\le 1$, the motion is bound, and thus a generalization
of the elliptical Kepler solutions. For $e'>1$, the motion is unbound, i.e.,
a generalization of hyperbolic Kepler solutions.
The boundary case $e'=1$ is bound at $\rho \to \infty$. The degenerate
case $e'=e''=0$ corresponds
to the limit $m\to 0$, $M <\infty$ \cite{TrumpSchieveBook}.
For the Type~II solutions, eccentricities in the range $e''>1$ produce
a pair of conjugate curves; the reduced trajectory follows one of the
two branches. Solutions in the range $e''\le 1$ are bound solutions
which are not differentiable at $\phi=(0,\pi)$ and thus
may be considered unphysical. Likewise the general solution
$e',e''\ne0$ appears to be unphysical on the same grounds based
on numerical investigation \cite{TrumpSchieveBook}.
The Type~I solutions 
were found by Cook \cite{Cook1972} who discussed only
the bound case. The bound Type~I orbits were also derived
by Piron and Reuse \cite{PironReuse1975}
by a transformation out of the center-of-mass rest frame to one in which
$\Delta t(\tau) \equiv 0$. 

The most interesting feature of the solutions is
the fact that
the perihelion of the bound two-body system does not
precess. This follows from the fact that the bound
orbit of any central potential is a closed curve \cite{TrumpSchieveBook}.
This is contrary to the 
prediction of a finite precession
made by Sommerfeld \cite{Sommerfeld1915}
using
the Heaviside-Lorentz equations in the one-body
limit, i.e., infinite mass of the source. 
This latter precession is entirely special relativistic, and in
the case of gravitation, the Sommerfeld rate
is exactly one-sixth the rate predicted by general relativity for the
same planetary masses. Sommerfeld proved that the nonvanishing
perihelion precession results
in the ``correct'' fine
structure hydrogen spectrum 
under the rule of semi-classical quantization \cite{Sommerfeld1915}.
This result has become
questionable, however, and it is generally regarded today 
as incorrect \cite{BetheSalpeter}. This judgment is based primarily on
the fact that
in the full quantum theory, fine structure is known to arise
intimately from the spin of the electron \cite{Pauli1927}, which has no
classical counterpart.
One should expect, therefore, that in the 
classical limit, the special relativistic two-body system
should not undergo perihelion precession.

It is interesting to impose the rule of semi-classical
quantization on the bound Type~I solutions above. In 
the case of the hydrogen atom, one should expect the result to
be
the Bohr prediction without
fine structure corrections. Since the energy spectrum is a
function of the separation of energy states, it is possible to
continue to ignore the energy of the center-of-mass and
to proceed by the reduced problem alone.

First it is necessary to solve the covariant Hamilton-Jacobi
equation to obtain the invariant action variables.
As in the Lagrangian version above, many of the 
results are valid generally for the conservative potential $V=V(\rho)$.
Following
Goldstein \cite{Goldstein}, ch.~10, it is useful to set aside
for the moment
the knowledge that the system can be reduced in dimensionality.
Using the  3+1-dimensional Lagrangian in (\ref{lagrangian}), 
the canonically conjugate 
momenta are
\begin{eqnarray}
p_{\rho}&=&m \dot{\rho},\quad
p_{\beta}=-m \rho \dot{\beta},\nonumber\\
\quad p_{\theta}&=& m \rho \cosh \beta \dot{\theta} ,
\quad p_{\phi}= m \rho \cosh \beta \sin{\theta} \dot{\phi}.
\label{momenta}
\end{eqnarray}
The Lorentz-invariant Hamiltonian is
\begin{equation}
K={1 \over {2m}}\left[{
p_{\rho}^2-{1 \over \rho}\left({
p_{\beta}^2 -
{1 \over{\cosh^2 \beta}}
\left[{
p_{\theta}^2 + {{p_\phi^2}\over {\sin^2 \theta}}}\right]
}\right) }\right]
+V(\rho).
\label{hamiltonian}
\end{equation}
\vfil\eject
Using the rule $p_{\rho}\to{{\partial S}/{\partial {\rho}}}$, etc.,
the 
energy equation $K=E$ becomes the {\em Hamilton-Jacobi equation},
\begin{eqnarray}
\left({ {{\partial S_{\rho}}\over{\partial \rho}} }\right)^2
&-&{1 \over {\rho^2}}
  \left[{
  \left({ {{\partial S_{\beta}}\over{\partial \beta}} }\right)^2
  -{1 \over {\cosh^2 \beta}}\cdot }\right.\nonumber\\
& &\cdot \left.{
   \left[{
    \left({ {{\partial S_{\theta}}\over{\partial \theta}} }\right)^2
    + {1 \over {\sin^2 \theta}} 
       \left({ {{\partial S_{\phi}}\over{\partial \phi}} }\right)^2
     }\right]\;
     }\right]
  +2mV(\rho)=2mE.
\label{hj}
\end{eqnarray}
Hamilton's characteristic function $S$
is
assumed here to be totally separable,
i.e.,
in the form $S=S_{\rho}+S_{\beta}+S_{\theta}+S_{\phi}$.
This allows one to solve
eq.~(\ref{hj}) by a procedure 
which is a generalization of the nonrelativistic method
for the
central force problem \cite{Goldstein}, i.e.,
using the invariant separation constants, 
$(\alpha_{\phi},\alpha_{\theta},\alpha_{\beta}, E)$ \cite{TrumpSchieveBook}.
This procedure 
yields the solutions
\begin{eqnarray}
{{\partial S}\over{\partial \phi}}&=&\alpha_{\phi},
\nonumber\\
& &\nonumber\\
{{\partial S}\over{\partial \theta}}&=&+\sqrt{\alpha_{\theta}^2
-{{\alpha_{\phi}^2}\over{\sin^2 \theta} }},
\nonumber\\
& &\label{hjsolutions}\\
{{\partial S}\over{\partial \beta}}&=&-\sqrt{
 {{\alpha_{\theta}^2}\over{\cosh^2\beta}}-{\alpha_\beta^2}},
\nonumber\\
& &\nonumber\\
{{\partial S}\over{\partial \rho}}&=&
+\sqrt{2mE-2mV-{{\alpha_{\beta}^2}\over{\rho^2}}}.
\nonumber
\end{eqnarray}
The action variables are given by 
$J_{\rho}=\oint{{{\partial S}/{\partial \rho}}\,d\rho}$, etc.,
where the integration is taken over the full
range of oscillation of the coordinate. 
The first two integrations are solved entirely in the manner
of the nonrelativistic problem, yielding
$J_\phi = 2\pi \alpha_\phi$
and
$J_\theta=2 \pi \left({ \alpha_\theta - \alpha_\phi }\right)$
The new step is the third integration (see \cite{TrumpSchieveBook}), 
which yields 
$J_\beta =2 \pi \left({ \alpha_\beta - \alpha_\theta }\right)$.
The last integration is over
libration of the radial coordinate $\rho$. The form of the integral
is identical to the corresponding nonrelativistic one for $r$, and thus
the identical method of contour integration is used \cite{Goldstein}, yielding
\begin{equation}
J_{\rho}=-\left({J_{\beta}+J_{\theta}+J_{\phi}}\right)
+{{2\pi\,i\,m\,k}\over{\sqrt{2mE}}}.
\end{equation}
It follows immediately that the energy is given by
\begin{equation}
K=-{{2\pi^2\,m^2\,k^2}
\over{\left({J_\rho+J_\beta+J_{\theta}+J_{\phi}}\right)^2}}=E.
\label{energy}
\end{equation}
The angular frequencies, which are given by the
rule $\nu_{\rho}=\partial K/\partial J_{\rho}$, etc., 
are therefore identical. Thus it is proved
what was asserted above,
namely that the bound orbits of the central force potential are closed.

The {\em rule of semi-classical quantization} stipulates that
$J_{\rho}=n_{\rho} h$, etc., 
where the quantum numbers $(n_{\rho}, n_{\beta}, n_{\theta},n_{\phi})$
take on integer values and where $h$ is Planck's constant.
From eq.~(\ref{energy}), the semi-classical energy levels in 
the case of the potential $V=-k/\rho$ 
are 
\begin{eqnarray}
E&=&-{{2\pi^2\,m\,k^2}\over{n^2\,h^2}}, \qquad n=1,2,3,\ldots,\\
n&=&n_{\rho}+n_{\beta}+n_{\phi}+n_{\theta}.
\label{E}
\end{eqnarray}
Assuming that $k$ represents the electrostatic interaction strength between
the proton and electron,
the prediction of the St\"{u}ckelberg special relativistic 
two-body theory gives the Bohr levels without fine structure corrections,
which is the correct
result in the case of no spin. 

\section{Comments on the Coordinate System}

The pseudospherical coordinates eqs.~(\ref{rho}) 
used in this paper
are those used by Cook \cite{Cook1972}
to study 
several two-body systems, 
including
the inverse-square potential $V=-k/\rho$. 
These coordinates
may be contrasted 
with
the alternate set of pseudospherical coordinates 
used
by Arshansky and Horwitz \cite{ArshanskyHorwitz1989} 
to solve 
the full quantum St\"{u}ckelberg model 
of the hydrogen atom
with no spin.  
Although Cook claimed to have found
in the full quantum case
an incorrect $n+1/2$ dependence of the hydrogen spectrum
on the principal
quantum number $n$, this result is obtained by
using
an electromagnetic retarded potential 
approximated for low particle acceleration 
(cf. eq.~(81) on p.~133 of ref.~(\cite{Cook1972}). 
The inverse-square
potential $V=-k/\rho$ studied here 
is discussed by Cook 
only in the purely classical
context, as a possible 
special relativistic gravitational solution. 
He used the bound two-body solutions 
he obtained for this potential
to study high-order
deviations from Newtonian motion 
in the Solar System orbits. 
The lack of perihelion precession discussed here is implicit in his work.

Regarding the lack 
of fine structure corrections
for the system without spin,
the semi-classical prediction obtained
in this paper 
is in agreement with the 
full quantum St\"{u}ckelberg result for this potential
obtained by Arshansky and Horwitz 
using the alternate pseudospherical coordinates. 
In the purely classical case,
the difference between the two coordinate systems is trivial,
since the two sets are identical 
in 2+1 dimensions. 
In the quantum case,
it may be possible 
to distinguish between the two coordinate systems
based on the support required for the reduced wave function (see 
ref.~\cite{ArshanskyHorwitz1989}).

\section{Acknowledgments}

We wish to acknowledge the helpful remarks of Prof.~Larry Horwitz 
of Tel Aviv University, particularly in regard to the question of coordinate
systems mentioned above.

\begin{figure}
\caption{ {\em Note for the electronic preprint version: this 
figure is viewable online as a GIF image at the address
http://order.ph.utexas.edu/mtrump/figures}
The classical orbits of the two-body problem
for Coulomb action-at-a-distance $V=-k/\rho$, $k>0$. The orbit
shows the separation between the two particle events in the
center-of-mass rest frame, using the Type~I and Type~II 
solutions in eqs.~(\ref{TypeI}) and~(\ref{TypeII}) as well as eq.~(\ref{phiorbit}).
The Type~I orbits correspond to an attractive interaction.
For $0<\left|{e'}\right|\le 1$, the system is bound, as in fig.~(a).
The bound orbit is a closed
curve which is differentiable; the nondifferentiability in
the diagram is due to numerical approximation.
For $\left|{e'}\right|>1$, the system is unbound, as in fig.~(b).
The Type~II solutions, for the case $e''>1$, 
correspond to repulsive scattering. The Type~II orbits for $e''<1$
are unphysical. It appears moreover that the general solution $e',e'' \ne 0$
is unphysical, based on numerical investigation. The orbit also
depends on the constant $f$, but variation of this constant, as well as variation
of $e',e''$ within a specified range above, results in a change of the
overall shape of the orbit but not in the type of motion.
[The values of the constants used in the examples are: 
(a) $e'=0.5$, $f=1.0$; (b) $e'=1.5$, $f=2.0$; (c) $e''=1.5$, $f=2.0$\,] }
\label{FigOne}
\end{figure}


\begin{thebibliography}{99}
\bibitem{Stueckelberg1941} E.~C.~G. St\"{u}ckelberg, 
     Helv.\ Phys.\ Acta {\bf 14},
     372, 588~(1941); {\bf 15}, 23~(1942).
\bibitem{HorwitzPiron1973} L.~P. Horwitz and C.~Piron, Helv.\ Phys.\ Acta
   {\bf 46}, 316~(1973).
\bibitem{Cook1972} J.~L. Cook, Aust.\ J.\ Phys.\ {\bf 25}, 117, 141 (1972).
\bibitem{PironReuse1975} C.~Piron and F.~Reuse, Helv.\ Phys.\ Acta
			{\bf 48}, 631~(1975).
\bibitem{TrumpSchieveBook} M.~A.\ Trump and W.~C.\ Schieve,
    {\em Classical Covariant Action-at-a-Distance}
    (Kluwer, Dordrecht, Neth.,to be published 1998); see also
     M.~A. Trump, Ph.~D dissertation, The Univ.\ of Texas at Austin, 1997.
\bibitem{TrumpSchieve1998} M.\ A.\ Trump and W.\ C.\ Schieve,
    ``Classical Scattering in an Invariant
      Coulomb Potential,'' Found.\ Phys., to be published 1998.
\bibitem{Sommerfeld1915} see A.~Sommerfeld, {\em Atomic Structure and 
     Spectral Lines, Vol.\ I }, (Methuen, London, 3rd ed.\ revised 1934;
     translated from the 5th German ed.\ by H.~L. Brose);
     see also J.~L. Synge, \begin{em}Relativity: The Special Theory\end{em} 
					(North-Holland, Amsterdam, 2nd ed.\ 1965).
\bibitem{Fanchi} J.~R. Fanchi, 
    {\em Parameterized Relativistic Quantum Theory}
    (Kluwer, Dordrecht, Neth., 1993).
\bibitem{HorwitzSchieve1992}L.\ P.\ Horwitz and W.~C.\ Schieve, 
   Phys.\ Rev.\ {\bf A 45}, 743 (1992).
\bibitem{TrumpSchieve1997}M.~A. Trump and W.~C. Schieve, Found.\ Phys.\
  {\bf 27} 1, 389~(1997).
\bibitem{Goldstein} H.~Goldstein, 
		\begin{em}Classical Mechanics\/\end{em} (Addison-Wesley, Reading, Mass.,
 2nd ed.~1980).
\bibitem{BetheSalpeter} H.~A. Bethe and E.~E. Salpeter, {\em The Quantum
    Mechanics of One- and Two-Electron
    Atoms} (Academic Press, N.~Y., 1957).
\bibitem{Pauli1927} W.~Pauli,  Z.\ Physik {\bf 43}, 601~(1927).
\bibitem{ArshanskyHorwitz1989} R. Arshansky and L. P. Horwitz, 
    J.\ Math.\ Phys.\  {\bf 30}, 66 (1989).
\end{thebibliography}
\end{document}